\documentclass[
,final            
  ]
  {aipproc}

\layoutstyle{8x11double}

\usepackage{amsfonts}
\usepackage{amsmath, amssymb}

\newcommand{\im}{\mathrm{im}}

\newcommand{\bcE}{{\check{E}}}

\newcommand{\bp}{\begin{Proposition}}
\newcommand{\ep}{\end{Proposition}}
\newcommand{\bl}{\begin{Lemma}}
\newcommand{\el}{\end{Lemma}}
\newcommand{\bt}{\begin{Theorem}}
\newcommand{\et}{\end{Theorem}}
\newcommand{\bd}{\begin{Definition}}
\newcommand{\ed}{\end{Definition}}
\newcommand{\End}{\mathrm{End}}

\newcommand{\g}{{\bf g}}

\newcommand{\eqdef}{\stackrel{{\rm def.}}{=}}

\newcommand{\cinf}{{{\cal \cC}^\infty(M,\R)}}

\DeclareFontFamily{U}{rsf}{}
\DeclareFontShape{U}{rsf}{m}{n}{<5> <6> rsfs5 <7> <8> <9> rsfs7 <10-> rsfs10}{}
\DeclareMathAlphabet\Scr{U}{rsf}{m}{n}

\newcommand{\KA}{K\"{a}hler-Atiyah~}

\def\hV{{\hat V}}

\def\Z{\mathbb{Z}}

\def\R{\mathbb{R}}

\def\rk{{\rm rk}}

\def\dd{\mathrm{d}}

\def\AdS{\mathrm{AdS}}

\def\cQ{{\cal Q}}

\newcommand{\be}{\begin{equation}}
\newcommand{\ee}{\end{equation}}
\newcommand{\ben}{\begin{equation}}
\newcommand{\een}{\end{equation}}
\newcommand{\beqa}{\begin{eqnarray*}}
\newcommand{\eeqa}{\end{eqnarray*}}
\newcommand{\beqan}{\begin{eqnarray}}
\newcommand{\eeqan}{\end{eqnarray}}

\newcommand{\nn}{\nonumber}

\newcommand{\id}{\mathrm{id}}

\newcommand{\tr}{\mathrm{tr}}

\def\per{\mathrm{per}}

\def\cC{{\mathcal C}}

\def\cB{\Scr B}

\def\cK{\mathrm{\cal K}}

\def\SO{\mathrm{SO}}

\def\cD{\mathcal{D}}

\def\cP{\mathcal{P}}
\def\cN{\mathcal{N}}

\def\cF{\mathcal{F}}
\def\cC{\mathcal{C}}

\def\G_2{\mathrm{G_2}}
\def\g2{\mathfrak{g}_2}

\def\Wh{\mathrm{Wh}}

\def\mb{\mathbf{b}}

\def\mf{\mathbf{f}}
\def\mF{\mathbf{F}}
\def\momega{{\boldsymbol{\omega}}}
\def\cQ{\check{Q}}
\def\f{\mathfrak{f}}

\begin{document}

\title{Foliated backgrounds for M-theory compactifications (I)}


\author{Elena Mirela Babalic}{
  address={Department of Theoretical Physics, National Institute of Physics and Nuclear Engineering\\
  Str. Reactorului no.30, P.O.BOX MG-6,  Bucharest--Magurele 077125, Romania, \\
   Department of Physics, University of Craiova, 13 Al. I. Cuza Str., Craiova 200585, Romania\\
\url{mbabalic@theory.nipne.ro}}
}

\author{Calin Iuliu Lazaroiu}{
  address={Center for Geometry and Physics, Institute for Basic Science (IBS), Pohang 790-784, 
  Republic of Korea\\
\url{calin@ibs.re.kr}}
}

\begin{abstract}
 We summarize our geometric and topological description of compact
 eight-manifolds which arise as internal spaces in ${\cal N}=1$ flux
 compactifications of M-theory down to $\AdS_3$, under the assumption
 that the internal part of the supersymmetry generator is everywhere
 non-chiral. Specifying such a supersymmetric background is {\em
 equivalent} with giving a certain codimension one foliation defined by a
 closed one-form and which carries a leafwise $G_2$ structure, a foliation whose
 topology and geometry we characterize rigorously.
 \end{abstract}

\maketitle


\section{Introduction}
Using the theory of foliations, we give a  mathematical characterization 
of the class of $\cN=1$ supersymmetric compactifications
of eleven-dimensional supergravity down to $\AdS_3$ spaces. Our results,
which rely on rigorous proofs, give a complete geometric and
topological characterization of those oriented, compact and connected
eight-manifolds $M$ which satisfy the corresponding supersymmetry
conditions, in the case when the internal part of the supersymmetry
generator is everywhere non-chiral. Details of the method, proofs and
calculations can be found in \cite{g2}.

An everywhere non-chiral Majorana spinor $\xi$ on $M$ can be parameterized by a
nowhere-vanishing 1-form $V$ whose kernel distribution $\cD$ carries
a $G_2$ structure. The condition that $\xi$ satisfies the
supersymmetry equations turns out to be {\em equivalent} with the requirements that
$\cD$ is Frobenius integrable (namely, a certain 1-form $\momega$ proportional
to $V$ must belong to a cohomology class specified by the supergravity
4-form field strength $\mathbf{G}$) and that the O'Neill-Gray
tensors of the codimension one foliation $\cF$ which integrates $\cD$,
the so-called ``non-adapted part'' of the normal connection of $\cF$
as well as the torsion classes of the $G_2$ structure of $\cD$ are
given in terms of $\mathbf{G}$ through explicit expressions. 
In particular, the leafwise $G_2$ structure is
``integrable'' (in the sense that its rank two torsion 
class $\boldsymbol{\tau}_2$ vanishes \cite{FriedrichIvanov2})
and conformally co-calibrated (the 
rank one torsion class $\boldsymbol{\tau}_1$ is exact), thus (up to a conformal
transformation) it is of the type studied in
\cite{GrigorianCoflow}. The field strength $\mathbf{G}$
can be determined in terms of the geometry of $\cF$ and
of the torsion classes of its longitudinal $G_2$ structure, provided
that $\cF$ and the $G_2$ structure satisfy some geometric
conditions, which we describe. We also describe the 
topology of $\cF$, thus providing 
a {\em global} description of $M$.

\paragraph{\bf Notations and conventions} 
Throughout the paper, $M$ denotes an oriented, connected and compact
smooth manifold of dimension 8, whose unital commutative $\R$-algebra of smooth real-valued functions we
denote by $\cinf$. All vector bundles considered are smooth. We use
 the results and notations of \cite{g2,ga1,gf}, with the same
conventions as there. For simplicity, we write the geometric
product $\diamond$ of loc. cit. simply as juxtaposition.
If $\cD\subset TM$ is a Frobenius distribution on $M$, we let
$\Omega(\cD)=\Gamma(M,\wedge \cD^\ast)$ denote the $\cinf$-module of
longitudinal differential forms along $\cD$.
For any 4-form $\omega\in \Omega^4(M)$, we let $\omega^\pm\eqdef
\frac{1}{2}(\omega\pm \ast \omega)$ denote the self-dual and
anti-selfdual parts (namely, $\ast \omega^\pm=\pm
\omega^\pm$).

\

As in \cite{Tsimpis, MartelliSparks}, we consider
supergravity on an eleven-dimensional connected and paracompact 
spin manifold $\mathbf{M}$ with Lorentzian metric $\mathbf{g}$ (of `mostly plus'
signature), compactified down to an $\AdS_3$ space of cosmological constant $\Lambda=-8\kappa^2$,
where $\kappa$ is a positive real parameter (this includes the
Minkowski case as the limit $\kappa\rightarrow 0$). 
The fields of the classical action are the metric $\mathbf{g}$, the
three-form potential $\mathbf{C}$ (with four-form field strength
$\mathbf{G}=\dd \mathbf{C}$) and the gravitino, which is a Majorana spinor of spin $3/2$.
 We can write $\mathbf{  M}=N\times M$ (a warped product), where $N$ is an oriented 3-manifold diffeomorphic
with $\R^3$ and carrying the $\AdS_3$ metric $g_3$, while $M$ is an oriented, 
compact and connected Riemannian 8-manifold with Riemannian metric $g$. 
 For the field strength $\mathbf{G}$, we use the warped compactification ansatz:
\ben
\label{Gansatz}
\mathbf{ G} = \nu_3\wedge \mathbf{f}+\mF~~,~~~~\mathrm{with}~~ 
\mF\eqdef e^{3\Delta}F~~,~~\mathbf{f}\eqdef e^{3\Delta} f~~\nn
\een
where $\nu_3$ is the volume form of $(N,g_3)$,  $f\in \Omega^1(M)$, $F\in \Omega^4(M)$ and 
$\Delta$ is the warp factor.

For supersymmetric bosonic classical backgrounds, the gravitino and
its supersymmetry variation must vanish, which requires the existence of
 at least one solution $\boldsymbol{\eta}$ to the supersymmetry condition:
\ben
\label{susy}
 \mathfrak{D} \boldsymbol{\eta} = 0~~,
\een
where $\mathfrak{D}$ denotes the supercovariant connection. We use the ansatz
$\boldsymbol{\eta}=e^{\frac{\Delta}{2}}(\zeta\otimes \xi)$ where
 $\xi$ is a Majorana spinor of spin $1/2$ on the internal space
$(M,g)$ (a section of the rank 16 real vector bundle $S$ of indefinite chirality real 
pinors) and $\zeta$ is a Majorana spinor on $(N,g_3)$. 
When $\zeta$ is a Killing spinor on $\AdS_3$, the
supersymmetry condition \eqref{susy} is equivalent with the following
system for $\xi$:
\ben
\label{par_eq}
\boxed{ \mathbb{D}\xi = 0~~,~~Q\xi = 0~}
\een 
where, for $X\in \Gamma(M,TM)$~:
\be
\mathbb{D}_X=\nabla_X^S+\frac{1}{4}\gamma(X\lrcorner F)
+\frac{1}{4}\gamma((X_\sharp\wedge f) \nu) +\kappa \gamma(X\lrcorner \nu)~\nn
\ee
is a linear connection on $S$  and 
\be
Q=\frac{1}{2}\gamma(\dd \Delta)-\frac{1}{6}\gamma(\iota_f\nu)
-\frac{1}{12}\gamma(F)-\kappa\gamma(\nu)~\nn
\ee 
is an endomorphism of $S$. Here $\nu$ is the volume form of $(M,g)$ and 
$\gamma:\wedge T^\ast M\rightarrow \End(S)$ is the structure morphism.
The set of solutions of \eqref{par_eq} is a finite-dimensional
subspace $\cK(\mathbb{D},{Q})$ of the infinite-dimensional
vector space of smooth sections of $S$. The
vector bundle $S$ has two admissible pairings $\cB_\pm$ (see \cite{gf,
  AC1}), both  symmetric but of different
 types $\epsilon_{\cB_\pm}=\pm 1$. Without loss of generality,
we choose to work with $\cB\eqdef \cB_+$ which can be taken
to be a scalar product on $S$ (see \cite{ga1}).

Requiring the background to preserve $\cN=1$ supersymmetry amounts to asking that
$\dim \cK(\mathbb{D},Q)=1$. Since $\cB$ is $\mathbb{D}$-flat (as proven in \cite{ga1}),
 any solution of \eqref{par_eq} of unit $\cB$-norm at a
point will have unit $\cB$-norm at every point of $M$ and thus we can take
the internal part $\xi$ of the supersymmetry generator to be
everywhere of norm one. 

\section{Geometric analysis}

\paragraph{\bf The inhomogeneous form defined by a Majorana spinor}
One can eliminate $\xi$ by working with differential forms constructed
as bilinears in $\xi$, a method which admits a concise mathematical
formulation through the theory of \KA bundles \cite{ga1,gf, ga2}.
Fixing a Majorana spinor $\xi\in \Gamma(M,S)$ which is of unit
$\cB$-norm everywhere, consider the inhomogeneus differential form
(see \cite{ga1}):
\ben
\label{checkE}
\check{E}=\frac{1}{16} \sum_{k=0}^8 \bcE^{(k)}=\frac{1}{16}(1+V+Y+Z+b\nu)~~,
\een
where we use the following notations for the non-zero components, 
which turn out to have ranks $k=0,1,4,5,8$ : 
\beqan
\label{forms8}
&&\bcE^{(0)}=||\xi||^2=1~~~,~~~\bcE^{(1)}\eqdef V   ~~,\nn \\
&& \bcE^{(4)}\eqdef Y ~~,~~\bcE^{(5)}\eqdef Z~~,~~\bcE^{(8)}\eqdef b\nu~~.\nn
\eeqan
The rank components of $\check{E}$ have the following expansions in any
local orthonormal coframe $(e^a)_{a=1\ldots 8}$ of $M$ defined on some
open subset $U\subset M$:
\ben
\label{Exi}
\bcE^{(k)}=_U\frac{1}{k!}\cB(\xi,\gamma_{a_1...a_k}\xi)e^{a_1...a_k} \in \Omega^k(M)~~.\nn
\een
Here, $b$ is a smooth real valued function defined on $M$. 
We have $b=||\xi^+||^2-||\xi^-||^2$ and
$||\xi^+||^2+||\xi^-||^2=||\xi||^2$.
The inequality $|b|\leq 1$ holds on $M$, with equality only at 
those $p\in M$ where $\xi_p$ has
definite chirality. The inhomogeneous differential form $\check{E}$ is 
the main object used in our method \cite{ga1,gf,ga2} based on \KA algebras.

\paragraph{\bf Restriction to Majorana spinors which are everywhere non-chiral}
From now on, we consider only the case when $\xi$ is everywhere
non-chiral on $M$, i.e. the case $|b|<1$ everywhere. This amounts to
requiring that each of the chiral components $\xi^\pm$ is
nowhere-vanishing. With this assumption, the Fierz identities (which we treated generally in \cite{gf,rec}) 
prove to be {\em equivalent} with the following relations holding 
on $M$\footnote{The first three of these relations were 
also given in \cite{MartelliSparks}.}:
\beqan
&& ||V||^2=1-b^2>0~~,\nn\\
&& \iota_V(\ast Z)=0~~,~~\iota_V Z=Y-b\ast Y~~,~~\label{SolMS}\\
&& (\iota_\alpha (\ast Z)) \wedge (\iota_\beta (\ast Z)) \wedge  (\ast Z) 
= - 6 \langle \alpha\wedge V, \beta\wedge V\rangle \iota_V \nu~,~\nn
\eeqan
for any 1-forms $ \alpha,\beta\in \Omega^1(M).$

The first relation in \eqref{SolMS} implies that $V$ is nowhere-vanishing. Thus $V$ determines a corank one Frobenius
distribution $\cD=\ker V\subset TM$ on $M$, whose rank one
orthocomplement (taken with respect to $g$) we denote by
$\cD^\perp$. This provides an orthogonal direct sum decomposition:
\be
TM=\cD\oplus \cD^\perp \nn
\ee
and  defines an orthogonal almost product structure $\cP\in
\Gamma(M,\End(TM))$. Equivalently, $V$ provides a reduction of structure group of $TM$ from $\SO(8)$ to
$\SO(7)$. For convenience, we introduce the normalized vector field:
\ben
\label{n}
n\eqdef \hV^\sharp=\frac{V^\sharp}{||V||}~~,~~||n||=1~~,\nn
\een
which is everywhere orthogonal to $\cD$ and generates
$\cD^\perp$. Thus $\cD^\perp$ is trivial as a real line bundle and
$\cD$ is transversely oriented by $n$. Since $M$ itself is oriented,
this provides an orientation of $\cD$ which agrees with that
defined by the longitudinal volume form
$\nu_\top=\iota_\hV\nu$ in the sense that~~$\hV\wedge \nu_\top=\nu$~.
We let $\ast_\perp:\Omega(\cD)\rightarrow \Omega(\cD)$ be the Hodge
operator along $\cD$, taken with respect to this orientation:
\ben
\label{AstV}
\ast_\perp \omega=\ast(\hV\wedge \omega)=(-1)^{\rk \omega}\iota_\hV(\ast \omega)~~,
~~\forall \omega\in \Omega(\cD)~,\nn
\een
where $\ast_\perp^2=\id_\Omega(\cD)$~ and $\ast^2=\pi$, where $\pi$
is the parity automorphism of the \KA algebra of $(M,g)$. Notice
that $\cD$ is endowed with the metric $g|_\cD$ induced by $g$, which,
together with the orientation defined above, gives $\cD$ an $\SO(7)$
structure as a vector bundle.

\paragraph{Proposition \cite{g2}} Relations \eqref{SolMS} are equivalent with:
\ben
\label{fsol}
V^2=1-b^2~~,~~Y=(1+b\nu)\psi~~,~~Z=V\psi~~,
\een
where $\psi\in \Omega^4(\cD)$ is the canonically normalized
coassociative four-form of a $G_2$ structure on the distribution $\cD$
which is compatible with the metric $g|_\cD$ induced by $g$ and with the
orientation of $\cD$ discussed above.

\paragraph{Remark} We remind the reader that the canonical normalization condition for
the coassociative 4-form $\psi$ (and thus also for the associative
3-form $\varphi\eqdef \ast_\perp \psi$) of a $G_2$ structure on $\cD$
is:
\ben
\label{CanNorm}
||\psi||^2=||\varphi||^2=7~~.\nn
\een
Since $\cD$ is a sub-bundle of $TM$, the Proposition  
shows that we have a $G_2$ structure which at every point $p\in M$ is given by the
isotropy subgroup $G_{2,p}$ of the pair $(V_p,\varphi_p)$ in the group
$\SO(8)_p\eqdef\SO(T_pM,g_p)$. Hence we have a two step reduction
along the inclusions: 
\be 
G_{2,p}\hookrightarrow \SO(7)_p\hookrightarrow \SO(8)_p~~, \nn
\ee 
where $\SO(7)_p\eqdef \SO(\cD_p,g_p|_{\cD_p})$ is the stabilizer of $V_p$ in
$\SO(8)_p$.

Notice that  $b,V,Y$ and $Z$ provide a redundant parameterization of $\xi$. 
A better parameterization is the one in terms of $b,V$ and $\psi$, where:
\ben
\label{YZinverse}
\psi=\frac{1}{1-b^2} VZ=\frac{1}{1-b^2}(1-b\nu)Y~~.\nn
\een
This parameterization is given by:
\ben
\label{Enr}
\boxed{\check{E}=\frac{1}{16}(1+V+b\nu)(1+\psi)=P\Pi}~ \nn
\een
where: 
\ben
\label{PiP}
P\eqdef \frac{1}{2}(1+V+b\nu)~~~\mathrm{and}~~~\Pi\eqdef \frac{1}{8}(1+\psi)  \nn
\een
are commuting idempotents in the \KA algebra of $(M,g)$. 
The fact that $P$ and $\Pi$
commute in this algebra is equivalent with the identity $\iota_V\psi=0$.
Idempotency of $P$ is equivalent with the relation $V^2=1-b^2$, while that of $\Pi$ is
equivalent with identity: 
\ben
\label{quadratic}
\psi^2=6\psi+7~,
\een
which follows from \eqref{SolMS}. It can be shown\footnote{We thank
  S. Grigorian for giving a direct proof of this statement.} that
condition \eqref{quadratic} characterizes metric-compatible
coassociative forms of $G_2$ structures on a Euclidean vector bundle
of rank $7$.

\subsection{Two problems related to the supersymmetry conditions}
\label{subsec:p}

We shall consider two different (but related) problems regarding
equations \eqref{par_eq}:
\paragraph{Problem 1} 
Given $f\in \Omega^1(M)$ and $F\in \Omega^4(M)$, find a set of
equations in $\Delta$ and $b,{\hat V},\psi$
  {\em  equivalent} with the supersymmetry conditions \eqref{par_eq}.
\paragraph{Problem 2} 
Find the necessary and sufficient {\em compatibility conditions} on
the quantities $\Delta$ and $b,{\hat V},\psi$ such that there exists
at least one pair $(f,F)\in \Omega^1(M)\times \Omega^4(M)$ for which
$\dim\cK(\mathbb{D},Q)>0$, i.e. such that \eqref{par_eq} admits at
least one non-trivial solution $\xi$.

\

In what follows, we describe only the solution of the second problem, 
following \cite{g2}; the solution of the first problem can also be
found in loc. cit.

\subsection{Encoding the supersymmetry conditions}

It was shown in \cite{ga1} that the supersymmetry conditions
\eqref{par_eq} are {\em equivalent} with the following equations for the
inhomogeneous form $\check{E}$, where commutators $[~,~]_-$ are taken in the \KA algebra
of $(M,g)$:
\ben
\label{NablaQConstr}
\boxed{
\begin{split}
&\nabla_m\check{E}=-[\check A_m,\check{E}]_{-}~\\
&\check{Q} \check E=0~,
\end{split}}
\een
the inhomogeneous forms $\check{A}_m$, $\check{Q}$ being given by:
\beqa
&& \check{A}_m=_U\frac{1}{4}e_m\lrcorner F+\frac{1}{4}(e^{m}\wedge f) \nu +\kappa e^{m}\nu~~~,\\~~~
&&\check{Q}=_U\frac{1}{2}\dd \Delta-\frac{1}{6}f\nu-\frac{1}{12}F-\kappa\nu~~.
\eeqa
We call the first row in \eqref{NablaQConstr} the {\em covariant derivative
  constraints}, while the second row describes the {\em
  $\cQ$-constraints}.
  
\paragraph{\bf Integrability of $\cD$. The foliations $\cF$ and $\cF^\perp$}
As already noticed in \cite{MartelliSparks}, the supersymmetry conditions
\eqref{par_eq} (equivalently, \eqref{NablaQConstr}) imply:
\ben
\label{dtV}
\boxed{
\begin{split}
&\dd \momega=0~~,~~\mathrm{where}~~\momega\eqdef 4\kappa e^{3\Delta} V~~,\\
&\momega=\mathbf{f}-\dd\mb~~,~~~\mathrm{where}~~\mb\eqdef e^{3\Delta} b~~.
\end{split}}
\een
In particular, the 1-form $\mf$ must be closed, a property which is also implied by
the Bianchi identities ($\dd \mathbf{G}=0$ leads to $\dd \mathbf{F}=\dd\mathbf{f}=0$). 
Relations \eqref{dtV} show that the closed form
$\momega$ belongs to the cohomology class of $\mf$. The first of these relations 
implies that the distribution $\cD=\ker V=\ker\momega$ is Frobenius integrable
and hence that it defines a codimension one foliation $\cF$ of $M$
such that $\cD=T\cF$. The complementary distribution is also
integrable (since it has rank one) and determines a foliation
$\cF^\perp$ such that $\cD^\perp=T(\cF^\perp)$. The leaves of
$\cF^\perp$ are the integral curves of the vector field $n$, which are
orthogonal to the leaves of $\cF$. The 3-form $\varphi$ defines a leafwise $G_2$ structure
on $\cF$. The restriction $S|_L$ of the pinor bundle $S$ to any given
leaf $L$ of $\cF$ becomes the bundle of real pinors of $L$, while the
restriction of a certain sub-bundle $S_+$ (described in \cite{g2}) becomes the bundle of Majorana spinors of the
leaf. Since the considerations of the present section are local, we can ignore for the
moment the global behavior of the leaves.
\begin{figure}[h!]
\includegraphics[scale=0.35]{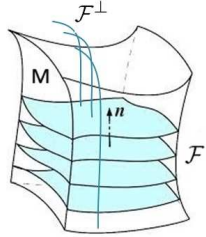}
\caption{Local picture of the plaques of the foliations $\cF$ and $\cF^\perp$ inside some open subset of $M$.}
\end{figure}

\paragraph{\bf Solving the $\check{Q}$ constraints}
Any form decomposes uniquely into components parallel and orthogonal 
to any 1-form. This gives
$F=F_\perp+\hV\wedge F_\top$ and
$f=f_\perp+\hV\wedge f_\top$, where $f_\top\in \Omega^0(M)$,
$f_\perp\in \Omega^1(\cD)$, $F_\top\in \Omega^3(\cD)$ and $F_\perp\in
\Omega^4(\cD)$. Since $\cF$ carries a leafwise
$G_2$ structure, $F_{\top,\perp}$ decompose according to the irreducible representations of $G_2$ into 
$F_{\top,\perp}=F^{(1)}+F^{(7)}_{\top,\perp}+F^{(27)}_{\top,\perp}=F^{(7)}_{\top,\perp}+F^{(S)}_{\top,\perp}$ and thus:
\be
F_\perp=F_\perp^{(7)}+F_\perp^{(S)}~~,~~F_\top=F_\top^{(7)}+F_\top^{(S)}  \nn
\ee
with the parameterization:
\beqan
\label{Fdecomp}
\boxed{
\begin{split}
& F_\perp^{(7)}=\alpha_1\wedge\varphi~~~,
~~~F_\perp^{(S)}=-\hat{h}_{kl}e^k\wedge\iota_{e^l}\psi~~\\
& F_\top^{(7)}=-\iota_{\alpha_2}\psi~~~,
~~~F_\top^{(S)}=\chi_{kl}e^k\wedge\iota_{e^l}\varphi~~.
\end{split}
}
\eeqan
Here $\alpha_1,\alpha_2\in\Omega^1(\cD)$ while ${\hat h}, \chi$ are
leafwise covariant symmetric tensors which decompose into their homothety parts $\tr_g({\hat h})$, $\tr_g(\chi)$ and
traceless parts $\chi^{(0)}$ and ${\hat h}^{(0)}$. 
Since $\psi=\ast_\perp\varphi$, relations \eqref{Fdecomp} determine $F$ in
terms of ${\hat V}$, $\psi$ and of the quantities
$\alpha_1,\alpha_2,{\hat h}$ and $\chi$. 
The tensor ${\hat h}$  has the properties (see \cite{Kthesis,Kflows}):
\beqan
{\hat h}_{ij}=\frac{1}{7}\tr_g({\hat h})g_{ij}+{\hat h}_{ij}^{(0)} &,&
~~\tr_g({\hat h}^{(0)})=0~~,\nn\\
{\hat h}_{ij}=h_{ij}-\frac{1}{4}\tr_g(h) g_{ij}
&\Longleftrightarrow & h_{ij}=\hat{h}_{ij}-\frac{1}{3}\tr_g(\hat h)g_{ij}~~,\nn
\eeqan
with similar properties for $\chi$.
The solution of the $\check{Q}$-constraints is given by the following:

\paragraph{Proposition \cite{g2}} 
Let $||V||=\sqrt{1-b^2}$. Then the $\cQ$-constraints 
are {\em equivalent} with the following relations, which determine (in
terms of $\Delta, b, {\hat V}, \psi$ and $f$) the components of
$F_\top^{(1)}$, $F_\perp^{(1)}$ and $F^{(7)}_\top$, $F_\perp^{(7)}$:
\beqan
\label{SolF}
\boxed{
\begin{split}
&\alpha_1=\frac{1}{2||V||}(f-3b\dd\Delta)_\perp  ~,\\
&\alpha_2= -\frac{1}{2||V||}(bf-3\dd\Delta)_\perp ~,\\
&\tr_g(\hat h)=-\frac{3}{4}\tr_g(h)=\frac{1}{2||V||}(bf-3\dd\Delta)_\top  ~,\\
&\tr_g(\hat \chi)=-\frac{3}{4}\tr_g(\chi)= 3\kappa- \frac{1}{2||V||}(f-3b\dd\Delta)_\top~.
\end{split}
}
\eeqan

\paragraph{Remark} 
Notice that the $\cQ$-constraints  do {\em not} determine the components $F_\top^{(27)}$ and $F_\perp^{(27)}$.

\paragraph{Definition} 
We say that a pair $(f,F)\in \Omega^1(M)\times \Omega^4(M)$ is {\em
  consistent} with a quadruple $(\Delta, b,{\hat V},\psi)$ if
conditions \eqref{SolF} hold, i.e. if the $\check{Q}$-constraints are
satisfied.

\paragraph{Encoding the covariant derivative constraints} In order to describe the extrinsic geometry of the foliation $\cF$
we start with the following form of {\em fundamental equations of the foliation}:
\beqan
\label{FE}
\boxed{
\begin{split}
&\nabla_n n=H ~~~(\perp n) ~,\\
&\nabla_{X_\perp} n=-A X_\perp ~~~(\perp n)~,\\
&\nabla_n (X_\perp)=-g(H,X_\perp) n+D_n(X_\perp)~,\\
&\nabla_{X_\perp} (Y_\perp)=\nabla^\perp_{X_\perp} (Y_\perp)+g(AX_\perp,Y_\perp)n~,
\end{split}~~
}
\eeqan
where $H\in \Gamma(M, \cD^\perp)$ encodes the second fundamental form
of $\cF^\perp$, ~$A\in \Gamma(M,\End(\cD))$ is the Weingarten operator
of the leaves of $\cF$ and $D_n:\Gamma(M,\cD)\rightarrow
\Gamma(M,\cD)$ is the derivative along the vector field $n$ taken with
respect to the normal connection of the leaves of $\cF^\perp$. The
first and third relations are the Gauss and Weingarten equations for
$\cF^\perp$ while the second and fourth relations are the Weingarten
and Gauss equations for $\cF$. It is shown in \cite{g2} that the following relations hold: 
\ben
\label{NormalPhiPsi}
\boxed{D_n\varphi=3\iota_\vartheta\psi~~~,~~~D_n\psi=-3\vartheta\wedge\varphi}~~,
\een
where $\vartheta\in \Omega^1(\cD)$.

\vspace{2mm}

We now give the solution of \emph{Problem 2}, which was 
obtained after lengthy calculation. The result gives the set of conditions equivalent with the
existence of at least one non-trivial solution $\xi$ of \eqref{par_eq}
which is everywhere non-chiral, while expressing $f$ and $F$ in terms
of $\Delta$ and of the quantities $b,{\hat V}$ and $\varphi$.

 \paragraph{\bf Theorem \cite{g2}} 
The following statements are equivalent: 
\begin{description}
\item{(A)} There exist $f\in \Omega^1(M)$ and $F\in \Omega^4(M)$ such that
  \eqref{par_eq} admits at least one non-trivial solution $\xi$ which
  is everywhere non-chiral (and which we can take to be everywhere of
  norm one).
\item{(B)} There exist $ \Delta\in \cinf$, $b\in \cC^\infty(M,(-1,1))$, 
  ${\hat V}\in \Omega^1(M)$ and $\varphi\in \Omega^3(M)$ such that:
\end{description}
\begin{enumerate}
\itemsep0em
\item $\Delta$, $b$, ${\hat V}$ and $\varphi$ satisfy the conditions:
\ben
\label{bVphi}
||{\hat V}||=1~~,~~\iota_{\hat V} \varphi=0~~.\nn
\een
The Frobenius distribution $\cD\eqdef \ker {\hat V}$ is
integrable and we let $\cF$ be the
foliation which integrates it.
\item The quantities $H$, $\tr A $ and $\vartheta$ of  $\cF$ are given by:
\beqan
\label{TgeomHB}
\boxed{
\begin{split}
&H_\sharp=\frac{2}{||V||}\alpha_2=-\frac{b}{||V||^2}(\dd b)_\perp+3(\dd\Delta)_\perp ~, \\
& \tr A =12(\dd\Delta)_\top-\frac{b(\dd b)_\top}{||V||^2}-8\kappa \frac{b}{||V||}~,\\
& \vartheta=-\frac{1+b^2}{6||V||^2}(\dd b)_\perp+\frac{b}{2}(\dd\Delta)_\perp~~.
\end{split}} \nn
\eeqan
\item $\varphi$ induces a leafwise $G_2$ structure on $\cF$ whose
  torsion classes satisfy (notice that $\boldsymbol{\tau}_3$ is not constrained):
\end{enumerate}
\vspace{-2mm}
\beqan
\label{TgeomTorsion}
\boxed{
\begin{split}
&\boldsymbol{\tau}_{0}=\frac{4}{7||V||}\Big[  2\kappa(3+ b^2)-\frac{3b}{2}||V||(\dd\Delta)_\top+\frac{1+b^2}{2||V||}(\dd b)_\top  \Big]\\
&\boldsymbol{\tau}_{1}= -\frac{3}{2}(\dd\Delta)_\perp~,\\
&\boldsymbol{\tau}_{2}=0~~.
\end{split}} \nn
\eeqan

\vspace{2mm}

\noindent In this case, the forms $f$ and $F$ are uniquely determined by
$b,\Delta, V$ and $\varphi$. Namely, the one-form $f$ is given by:
\ben
\label{Tf}
\boxed{f=4\kappa V+e^{-3\Delta}\dd(e^{3\Delta} b)}~~,\nn
\een
while $F$ is given as follows:
\begin{description}
\item{(a)}  $F^{(1)}_\top=-\frac{4}{7}\tr_g(\hat\chi)\varphi$~~ and~~ $
  F^{(1)}_\perp=-\frac{4}{7}\tr_g(\hat h)\psi$, ~~with:
\ben
\label{TF1}
\boxed{
\begin{split}
&\tr_g(\hat h)=-\frac{3||V||}{2}(\dd\Delta)_\top+2\kappa b+
\frac{b}{2||V||}(\dd b)_\top\\
&\tr_g(\hat\chi)=\kappa-\frac{1}{2||V||}(\dd b)_\top
\end{split}\nn
}
\een
\item{(b)}  $F^{(7)}_\top=-\iota_{\alpha_2}\psi$~~ and
  ~~$F^{(7)}_\perp=\alpha_1\wedge\varphi$,~~ with:
\ben
\label{TF7}
\boxed{
\begin{split}
&\alpha_1=\frac{1}{2||V||}(\dd b)_\perp~~\\
&\alpha_2=-\frac{b}{2||V||}(\dd b)_\perp+\frac{3||V||}{2}(\dd\Delta)_\perp
\end{split}} \nn
\een
\item{(c)} $ F^{(27)}_\perp=-h^{(0)}_{ij} e^i\wedge\iota_{e^j}\psi $~~,~~$ F^{(27)}_\top=\chi^{(0)}_{ij} e^i\wedge\iota_{e^j}\varphi$,~with:
\end{description}
\ben
\boxed{
\begin{split}
h^{(0)}_{ij}=-\frac{b}{4||V||}[\langle e_i\lrcorner \varphi, e_j\lrcorner\boldsymbol{\tau}_{3}\rangle + (i\leftrightarrow j)]-\frac{1}{||V||}A^{(0)}_{ij}\\
\chi^{(0)}_{ij}=-\frac{1}{4||V||}[\langle e_i\lrcorner \varphi, e_j\lrcorner\boldsymbol{\tau}_{3} \rangle+ (i\leftrightarrow j)] - \frac{b}{||V||}A^{(0)}_{ij}
\end{split}} \nn
\een
where $A^{(0)}=\frac{1}{||V||}(b \chi^{(0)}_{ij}-h^{(0)}_{ij})$ is the traceless part of the Weingarten tensor of $\cF$ while $\boldsymbol{\tau}_3$ is the rank 3 torsion class of the leafwise $G_2$
structure.

The torsion classes $\boldsymbol{\tau}_0\in\Omega^0_1(L)$,
$\boldsymbol{\tau}_1\in \Omega^1_7(L)$, $\boldsymbol{\tau}_2\in
\Omega^2_{14}(L)$ and $\boldsymbol{\tau}_3\in \Omega^3_{27}(L)$ of the
$G_2$ structure are uniquely specified through the following equations,
which follow the conventions of \cite{Kthesis, Kflows, Bryant}:
\beqan
\label{TorsionEqs}
\begin{split}
&\dd_\perp\varphi=\boldsymbol{\tau}_0\psi+3\boldsymbol{\tau}_1\wedge\varphi+\ast_\perp\boldsymbol{\tau}_3~~,\\
&\dd_\perp\psi=4\boldsymbol{\tau}_1\wedge\psi+\ast_\perp\boldsymbol{\tau}_2~~.\nn\\
\end{split}
\eeqan
In our case, the leafwise $G_2$ structure is ``integrable'' (since 
$\boldsymbol{\tau}_2=0$), i.e. it belongs to the class $W_1\oplus
W_3\oplus W_4$ of the Fernandez-Gray classification. It is also conformally co-calibrated, 
since $\boldsymbol{\tau}_1$ is exact along each leaf. 

\subsection{Topological obstructions}

The condition that $\xi$ be nowhere chiral imposes constraints on the cohomology class 
$\f$ of $\mf$ and on the topology of $M$. A necessary condition is that 
$\f$ must be a non-trivial cohomology class and hence the first Betti number of $M$ cannot be zero:
\ben
\label{Betti}
\boxed{b_1(M)>0}~~.
\een
This condition is far from sufficient. 
The necessary and sufficient conditions are as follows. Let
$\hat{M}_\f$ be the integration cover of the period map $\per_\f$ of
$\f$, i.e. the Abelian regular covering space of $M$ corresponding to
the normal subgroup $\im (\per_\f)$ (known as the 
{\em group of periods} of $\f$) of $\pi_1(M)$. Then the class $\f \in
H^1(M,\R)\setminus \{0\}$ contains a nowhere-vanishing closed one-form
iff. $M$ is $(\pm \f)$-contractible and the so-called {\em Latour
obstruction} $\tau_L(M,\f)\in \Wh(\pi_1(M),\f)$ vanishes. Here
$\Wh(\pi_1(M),\f)$ is the Whitehead group of the Novikov-Sikorav ring
$\widehat{\Z\pi_1(M)}$. When $\f$ is projectively rational, these
conditions are equivalent with the requirements that the integration
cover $\hat{M}_\f$ (which in that case is infinite cyclic) must be
finitely-dominated and that the {\em Farrell-Siebenmann obstruction}
$\tau_F(M, \f)\in \Wh(\pi_1(M))$ vanishes, where $\Wh(\pi_1(M))$ is
the Whitehead group of $\pi_1(M)$. We refer the reader to \cite{g2} and to the 
references therein for more information.

\subsection{Topology of the foliation}

We briefly summarize some aspects of the topology of $\cF$, referring
the reader to \cite{g2} for details. We already noticed that $\cF$ is
transversely orientable. This implies that the holonomy group of each
leaf of $\cF$ is trivial, that all leaves of $\cF$ are diffeomorphic
with each other and that $\pi_1(L)$ can be identified with the group
of periods of $\f$. The character of the foliation depends on the rank
$\rho$ of the period group.

\paragraph{When $\momega$ is projectively rational, i.e. $\rho=1$} 
In this case, the leaves of $\cF$ are compact and coincide with the
fibers of a fibration $\mathfrak{h}:M\rightarrow S^1$.  Moreover, $M$
is diffeomorphic to the mapping torus
$\mathbb{T}_{\phi_{a_\f}}(M)\eqdef M\times[0,1]/\{(x,0)\sim
(\phi_{a_\f}(x),1)\}$, where $a_\f$ is the fundamental period of $\f$.

\paragraph{When $\momega$ is projectively irrational, i.e. $\rho>1$}
In this case, each leaf of $\cF$ is non-compact and dense in $M$ and
hence $\cF$ cannot be a fibration. The quotient topology on the space
of leaves is the coarse topology. In this case, we say that $\cF$ is a 
{\em minimal} foliation. 

\vspace{0.2cm}

The case when $\cF$ is minimal is generic. In that case, our
compactifications {\em cannot} be interpreted as ``generalized
Scherk-Schwarz compactifications with a twist'' (see \cite{Vandoren}) 
of some effective four-dimensional theory.

\paragraph{Noncommutative topology of the leaf space} 

Let $C(M/\cF)$ be the $C^\ast$-algebra of the foliation, which encodes
the `noncommutative topology' of its leaf space in the sense of
A.~Connes \cite{ConnesNG}.  Since the leaves of $\cF$ have no
holonomy, the explicit form of this $C^\ast$-algebra can be determined
up to Morita equivalence.  Let $\Pi_\f\approx \Z^\rho$ be the group of
periods of $\f$.  Then $C(M/\cF)$ is separable and strongly Morita
equivalent (hence also stably isomorphic) with the crossed product
algebra $C_0(\R)\rtimes \Pi_\f$, which is isomorphic with $C(S^1)$
when $\rho=1$ and with a $\rho$-dimensional noncommutative torus when
$\rho>1$. It follows that linear foliations of tori are so-called
``model foliations'' for $\cF$.
\vspace{-2mm}
\begin{figure}[h!]
\includegraphics[scale=0.31]{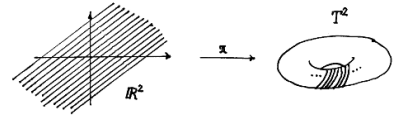}
\caption{The linear foliations of $T^2$ model the noncommutative geometry of the leaf space of $\cF$ in the case $\rho(\f)\leq 2$.}
\end{figure}


\begin{theacknowledgments}
E.M.B. acknowledges funding from the 
  strategic grant POSDRU/159/1.5/S/133255, Project ID 133255 (2014),
  co-financed by the European Social Fund within the Sectorial
  Operational Program Human Resources Development 2007--2013, 
while the work of C.I.L. is supported by the research  grant IBS-R003-G1. 
This work was also financed by the CNCS-UEFISCDI grants PN-II-ID-PCE 121/2011 
and 50/2011, and by PN 09 370102. 
\end{theacknowledgments}



   \bibliographystyle{aipproc}   



\end{document}